\documentclass[twocolumn,aps,pra,groupedaddress,showpacs]{revtex4}
\usepackage{epsfig,amssymb}

\def\bquar{\rlap{\hbox{$\swarrow$}}\hbox{$\nearrow$}}
\def\bmquar{\rlap{\hbox{$\nwarrow$}}\hbox{$\searrow$}}
\def\quar{{\rlap{\hbox{\tiny{$\swarrow$}}}\hbox{\tiny{$\nearrow$}}}}
\def\mquar{{\rlap{\hbox{\tiny{$\nwarrow$}}}\hbox{\tiny{$\searrow$}}}}
\begin{document}
  \title{Conveyor belt clock synchronization}
\author{Vittorio Giovannetti,$^1$\footnote{Now with NEST-INFM \& Scuola
Normale Superiore, Piazza dei Cavalieri 7, I-56126, Pisa, Italy.} Seth
Lloyd,$^{1,2}$ Lorenzo Maccone,$^1$\footnote{Now with QUIT - Quantum 
Information Theory Group
Dipartimento di Fisica ``A. Volta''
Universita' di Pavia, via A. Bassi 6
I-27100, Pavia, Italy.} Jeffrey H.  Shapiro,$^1$ and Franco
N. C. Wong$^1$}
\affiliation{Massachusetts Institute of
   Technology, \\$^1$Research Laboratory of Electronics,\\
   $^2$Department of Mechanical Engineering,\\ 77 Massachusetts Ave.,
   Cambridge MA 02139, USA.}  \date{\today}
\begin{abstract}
   A protocol for synchronizing distant clocks is proposed that does not
rely on the arrival times of the signals which are exchanged, and an
optical implementation based on coherent-state pulses is described.  This
protocol is not limited by any dispersion that may be present in the
propagation medium through which the light signals are exchanged.
Possible improvements deriving from the use of quantum-mechanical
effects are also addressed.
\end{abstract}
\maketitle

The synchronization of distant clocks is of considerable
importance for communications, multi-processor computations,
astronomy, geology, the global positioning system (GPS), etc.  Existing
synchronization protocols fall into two categories: Eddington adiabatic
transfer {\cite{eddi}} and Einstein clock synchronization {\cite{einst}}.
Eddington's method requires that the two parties (say Alice and Bob)
exchange a running clock, e.g., Alice sends her clock to Bob,
and he  compares it with his own.  This method does not require
time-of-arrival measurements, but it is usually impractical because a
complex system (a clock) must be exchanged. It is much
easier to implement Einstein's method, in which all that is exchanged is
a sequence of signal pulses, e.g., Alice sends
a signal pulse to Bob, which he then returns to Alice.  By recording the
signal's times of departure and arrival, Alice and Bob can synchronize
their clocks.  A variation of one or the other of these protocols is
invariably employed whenever two clocks must be synchronized
{\cite{review}}: either it is necessary to exchange clocks, or there is
an explicit dependence on time-of-arrival measurements. Typical
examples of Einstein clock synchronization are the ``two way'' protocols
in which Alice and Bob both exchange signals, phase-locked loop
techniques, and pseudo-random code correlation measurements such as
are used in GPS.

Here we discuss a synchronization protocol that is neither equivalent
to Eddington nor to Einstein synchronization, but instead embodies
the best features of each.  As in Einstein's scheme, it
is based on exchanging signals, thus avoiding the technological
problems associated with the exchange of complex systems such as
clocks (``shocks on clocks'') or entangled systems {\cite{jozsa}}. As in
Eddington's scheme, no time-of-arrival measurements are required, thus
avoiding the problems associated with such measurements, {e.g.}, those
arising from  dispersion in the signal's propagation medium.
In this paper we will focus on implementations that rely on
classical signals, but the method is well-suited for intrinsically
quantum-mechanical clock synchronization protocols {\cite{prl}}.

In Sec.~{\ref{s:schema}} we introduce the ``conveyor belt'' protocol
and describe its basic features (some useful variations are discussed in
App.~A).  A list of possible implementations in different physical
contexts is also given.  In Sec.~{\ref{s:sasha}} we present an
implementation that relies on polarized laser pulses.  Under rather
general conditions it is shown that this implementation's
attainable synchronization accuracy is
unaffected by any dispersion which may be present in the propagation
medium.  In Sec.~{\ref{s:quantum}} we show how quantum-mechanical effects
may be used to enhance the protocol's dispersion suppression:  employing
frequency-entangled pulses affords dispersion cancellation in
even more general circumstances than is the case for
implementations using classical (laser) light pulses.

\section{``Time independent'' clock synchronization}\label{s:schema}
In this section we describe in detail the conveyor belt
synchronization scheme, which was first proposed in {\cite{prl}}. The
two pre-conditions that must be satisfied are those underlying
Einstein's protocol: a) we need a physical medium that supports
signaling between Alice and Bob in which the Alice-to-Bob and
Bob-to-Alice transit times, $T_{ab}$ and $T_{ba}$, are identical;
b) we require Alice and Bob to have near-perfect, albeit unsynchronized,
clocks, viz., their relative drift is negligible over a roundtrip time
$2T$, where $T = T_{ab}=T_{ba}$. [In App.~A we discuss some
variations of our scheme which permit some softening of these
requirements].

\begin{figure}[hbt]
\begin{center}\epsfxsize=.9
\hsize\leavevmode\epsffile{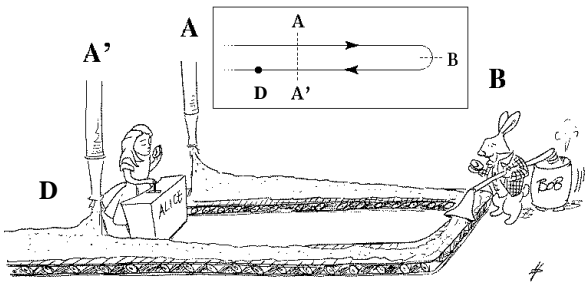}
\end{center}
\caption{Pictorial representation of the conveyor belt synchronization
scheme. Alice pours sand on the conveyor belt at positions $A$ and $A'$,
while Bob scoops away sand at the intermediate position $B$.
Measuring the amount of sand at position $D$---once an initial
transient has passed---directly reveals the time difference between their
two clocks.}
\label{f:nastro}\end{figure}

Our protocol can be explained by means of a simple illustrative
scenario. Suppose that there is a conveyor belt connecting Alice
and Bob, as shown in Fig.~{\ref{f:nastro}}, moving at speed $\nu$.
Upon initiation of the protocol, and continuing until its completion,
Alice pours sand onto the belt at points $A$ and $A'$ according to the
following schedule:  when her clock reads $t^a$ she deposits sand at
rate
$st^a/2$ at both $A$ and
$A'$.  Bob, for his part, removes sand at rate $st^b$ from
point $B$ when his clock reads $t^b$.
Alice completes the protocol by monitoring the amount of sand at 
point $D$---which is
after point $A'$ on the conveyor belt---as a function of
$t^a$, and waiting for it to stabilize to a constant value $Q_D$.  It is
easy to see that $Q_D$ is proportional to the time difference between
Alice's clock and Bob's clock, as we now demonstrate.

In terms of an external reference clock, showing time $t$, we may express
$t^a$ and $t^b$---the times shown on the clocks in Alice's and Bob's
possession---as follows:
\begin{equation}
t^a = t -t_0^a\quad\mbox{and}\quad t^b = t - t_0^b.
\end{equation}
Here, $t_0^b - t_0^a$ is the offset between Alice's clock and
Bob's that the conveyor belt protocol is trying to measure.  Once the
initial transient is over, i.e., when
$t\ge
\max(2T +t_0^a,t+t_0^b)$, we find that
\begin{eqnarray}
\!\!\!Q_D\!\!&=&\!\!\frac {s}{2}(t-2T-t_0^a)-s(t-T-t_0^b)+
\frac{s}{2}(t-t_0^a) \label{sabbia}\\&=&\!\!s(t_0^b-t_0^a)
\;\label{sabbia1},
\end{eqnarray}
where the first term on the right-hand side of Eq.~(\ref{sabbia}) is the
amount of sand that Alice deposited at point $A$ at time $t-2T$,
the second term is the amount of sand that Bob removed from point
$B$ at time $t-T$, and the third term is the amount of sand that
Alice added at position $A'$ at time $t$.

The three main features of this scheme are: 1) no time measurements are
needed; 2) the only role played by the signal transit time between
Alice and Bob is setting the duration of the transient
that must be endured before the synchronization measurement can be made;
and 3) the synchronization precision only depends on the precision with
which sand may be added to, removed from, and measured on the
conveyor belt.

That our protocol differs dramatically from Einstein
synchronization can be seen from the fact that ours is transit-time
independent, i.e., except for its impact on the duration of the
conveyor-belt transient, the transit time $T$---hence the distance
between Alice and Bob, $L = \nu T$---plays no role in the
protocol.  Indeed, neither Alice nor Bob need to know $T$ to run the 
protocol, nor
can they to deduce this transit time by measuring the post-transient 
amount of sand
on the belt at point
$D$.   A simple modification of our scheme, however, does permit $T$ to
be measured, so that the distance between Alice and Bob may be 
inferred if $\nu$ is known:
Alice continues to add sand at rate $st^a$ at point
$A$, Bob ceases any action at point $B$, and Alice removes sand at
rate $st^a$ from point $A'$.  Once the ensuing transient is over,
the amount of sand on the conveyor belt at point $D$ will be
\begin{eqnarray}
\!\!\!Q_D\!\!&=&\!\!\frac {s}{2}(t-2T-t_0^a)-
\frac{s}{2}(t-t_0^a) = -sT
\;.
\end{eqnarray}
If we use microwave signal propagation in lieu of a conveyor belt and 
the imposition of a
positive (negative) frequency shift instead of adding (removing) 
sand, the ranging
protocol we have just described is then the familiar 
frequency-modulated continuous wave (FMCW)
radar {\cite{radar}}.

Now, having illustrated the essentials of conveyor belt
clock synchronization in terms of the sand-based protocol, let us
address more realistic implementations.   Alice and Bob may exchange
electrical signals, whose voltages are modulated in accord with the
conveyor belt idea.  Alternatively, they may transmit sound waves (as in
sonar applications), modulating their frequencies to achieve clock
synchronization via our protocol.  The most appealing scenario, however,
involves light pulses.  In this context Alice and Bob may
encode synchronization information on the pulses using the polarization
direction (through Faraday rotators), frequency (through acousto-optic
modulators) or phase (through electro-optic modulators).  An application
of this type is analyzed in the next section.

\section{Dispersion-immune synchronization}\label{s:sasha}
Dispersion-induced pulse spreading and pulse distortion are
among the principal performance-limiting factors in
schemes that are currently used to synchronize distant clocks
{\cite{review}}.  We can exploit our protocol's independence of
time-of-arrival measurements to devise synchronization schemes that
thwart the ill-effects of dispersion.  In a
previous paper {\cite{prl}} we achieved this goal by means of
quantum-mechanical effects. Here, we show that classical pulses can be
used to achieve similar dispersion immunity under a wide range of
conditions.

\begin{figure}[hbt]
\begin{center}\epsfxsize=.9
\hsize\leavevmode\epsffile{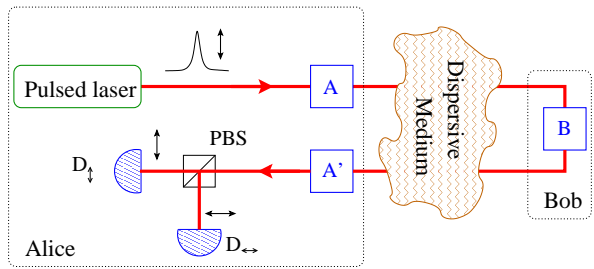}
\end{center}
\caption{Proposal for dispersion-immune synchronization.
The laser produces intense
$\updownarrow$-polarized pulses that travel from Alice to Bob, where they
are reflected back to Alice.  At points $A$ and $A'$, Alice delays the
$45^\circ$ polarization with respect to the $-45^\circ$ polarization by
an amount proportional to the time shown on her clock.  At point $B$, Bob
delays the $-45^\circ$ polarization with respect to the $45^\circ$
polarization by an amount proportional to the time shown on his clock.  These
delays are in accord with the conveyor belt protocol, i.e., Bob's
proportionality constant is twice Alice's.
The polarizing beam splitter PBS
separates the incoming beam into its $\updownarrow$ and $\leftrightarrow$
polarization components.  These components are directed to
integrating detectors $D_\updownarrow$ and $D_\leftrightarrow$
respectively, which measure the number of photons impinging on them. 
As discussed
in the text, signal multiplexers allow pulses to travel
through the dispersive medium in a common polarization state, thus
avoiding polarization-dependent propagation effects.  }
\label{f:schema}\end{figure}

The configuration for classical light-pulse clock synchronization via the
conveyor belt protocol is shown in Fig.~{\ref{f:schema}}.  In essence, it
is a polarization-based, time-delay interferometer.  A linearly
polarized (say $\updownarrow$) laser source emits intense light pulses of
center frequency $\omega_0$ and bandwidth
$\Delta\omega$.  Conveyor belt encoding and decoding is achieved by means
of time delays.  In particular:  at points $A$ and $A'$, Alice delays the
$45^\circ$ ($\bquar$) polarization with respect to the $-45^\circ$
($\bmquar$) polarization by an amount proportional to the time shown on her
clock; and at point $B$, Bob delays the $-45^\circ$ polarization with
respect to the
$45^\circ$ polarization by an amount proportional to the time shown on his
clock, with Bob's proportionality constant being twice Alice's.
The net effect of these actions, as seen at the input port to the
polarizing beam splitter PBS, is to delay the $\bquar$ component of the
returning light pulse relative to that pulse's $\bmquar$
component by $\tau_D = \beta(t_0^b-t_0^a)$, where $\beta$ is Bob's
proportionality constant (a dimensionless quantity).  Alice now obtains
the desired synchronization information by measuring $J_\leftrightarrow$,
the average photon number in the horizontally-polarized component of the
return pulse, by means of the polarizing beam splitter and the
integrating photodetector $D_\leftrightarrow$.  Because no time-of-arrival
information is sought in this measurement, dispersion can be neglected if
the $\bquar$ component encounters the same dispersion as its $\bmquar$
counterpart. As shown below, where we analyze the behavior of
the Fig.~{\ref{f:schema}} system, this common-mode dispersion condition
can be relaxed in several ways.

\begin{figure}[hbt]
\begin{center}\epsfxsize=.75
\hsize\leavevmode\epsffile{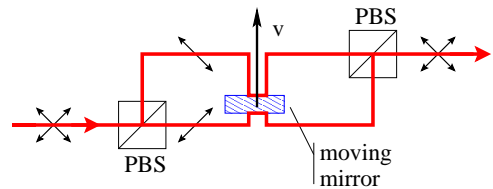}
\end{center}
\caption{Model of the time-varying delays introduced by Alice at points
$A$ and $A'$ in the Fig.~\ref{f:schema} system.  The left polarizing beam
splitter (PBS) separates the two polarization components so that they
impinge on opposing faces of a mirror moving at speed $v$ in the
direction shown. The right PBS recombines the polarizations.  Bob uses a
similar setup at point $B$ in the Fig.~\ref{f:schema} system, but his mirror
moves at speed $2v$ in the opposite direction from what is shown here.
Electro-optic modulators would be used, instead of the moving mirror, in
an actual system.}
\label{f:delay}\end{figure}

Before delving into the mathematics, an initial comment about our
theoretical approach is warranted.  We will employ quantum photodetection
theory in our treatment, despite that fact that semiclassical
(shot-noise) theory is quantitatively correct for the
Fig.~{\ref{f:schema}} system because it uses coherent-state (classical)
light \cite{PartIII}.  Our choice in this regard makes it more
difficult to connect our work to the literature on laser radar
\cite{osche}, which relies on semiclassical theory and could be used,
e.g., to address the performance of time-of-arrival measurements for
light-pulse Einstein synchronization.  Our reason for choosing to use
quantum theory is to enable an easy transition to assessing the additional
benefits that accrue from the use of nonclassical light---specifically
entangled states---in conveyor belt synchronization.  Semiclassical
photodetection is unable to treat such systems correctly.

The average photon flux arriving at detector D$_\leftrightarrow$ at time
$t$ is given by {\cite{mandel}}
\begin{eqnarray} I_\leftrightarrow(t)=\langle \Psi|
\;E^{(-)}_\leftrightarrow (t) E^{(+)}_\leftrightarrow (t)\;|\Psi\rangle
\;\label{intens1},
\end{eqnarray}
where $|\Psi\rangle$ is the quantum state of the light emitted by the
source and the field operators at the detector are given by
\begin{eqnarray}
E^{(+)}_\leftrightarrow (t)=
\left(E^{(-)}_\leftrightarrow(t)\right)^\dag=
\int
{\rm d}\omega
\;A_\leftrightarrow(\omega)\; e^{-i\omega
t}\;\label{campo}.
\end{eqnarray}
The annihilation operator $A_\leftrightarrow(\omega)$ destroys a
$\leftrightarrow$ polarized photon of frequency $\omega$ at the location
of detector $D_\leftrightarrow$. The average photon flux arriving at
detector $D_\updownarrow$ is obtained in a similar manner. In order to
connect the operators $A_\leftrightarrow(\omega)$ and
$A_\updownarrow(\omega)$ with those at the source, we first
express the
$\updownarrow$ and $\leftrightarrow$ components in terms of their
$\bquar$ and $\bmquar$ counterparts:
\begin{eqnarray}
A_\leftrightarrow(\omega)&=&\frac
1{\sqrt{2}}\left(A_\quar(\omega)-A_\mquar(\omega)\right)\;\;\;
\label{trasf1}\\
A_\updownarrow(\omega)&=&\frac
1{\sqrt{2}}\left(A_\quar(\omega)+A_\mquar(\omega)\right)\;.
\;\label{trasf2}
\end{eqnarray}
The annihilation operators $A_\quar$ and $A_\mquar$ may be now linked to
the corresponding annihilation operators $a_\quar$ and $a_\mquar$ at the
source position by accounting for the time-varying
delays that Alice and Bob impose in the conveyor belt protocol.
Their actions are equivalent to what occurs in the
Fig.~{\ref{f:delay}} arrangement, in which the two polarizations impinge
on opposite faces of a moving mirror.
Electro-optic modulators would be employed in an actual application, 
but the idealized
Fig.~{\ref{f:delay}} setup affords us an easy route to calculating the
field evolution from the source to the detector.

Alice has two Fig.~{\ref{f:delay}} setups, one at point $A$ and one
at point $A'$.  At time $t_0^a$ she starts moving both of her
mirrors  with constant speed $v$, imparting a Doppler frequency
shift---in the non-relativistic, $v\ll c$, limit---$v\omega/c$
($-v\omega/c$) to the $\bmquar$ ($\bquar$) polarization of an incoming
frequency-$\omega$ field, where $c$ is the phase velocity in the
propagation medium at frequency $\omega$. Bob, on the other hand, starts
moving his mirror---located at position
$B$---at time
$t_0^b$ with constant speed $2v$ in the opposite direction to what Alice
employs.  Thus, his action leads to a Doppler frequency shift
$-2v\omega/c$ ($2v\omega/c$) on the $\bmquar$ ($\bquar$)
polarization of an incoming frequency-$\omega$ field.
It follows that the overall annihilation operator transformation that we
are after is
\begin{eqnarray}
a_\quar(\omega)\;&\longrightarrow&\;A_\quar(\omega)=a_\quar(\omega)
\;e^{-i\omega\tau_D+i\omega\tau+i\kappa_\quar(\omega)}\label{delays1}
\\
a_\mquar(\omega)\;&\longrightarrow&\;A_\mquar(\omega)=a_\mquar(\omega)
\;e^{i\omega\tau_D+i\omega\tau+i\kappa_\mquar(\omega)}
\;\label{delays2},
\end{eqnarray}
where the term $\tau\equiv 2L/c$ accounts for the distance $L$ separating
Alice and Bob, and
\begin{eqnarray}
\tau_D&\equiv&-{4v}(t_0^b-t_0^a)/c
\;\label{deftheta}
\end{eqnarray}
contains the time shift that is needed to synchronize Alice's clock with
Bob's.  Note that we have neglected propagation loss in the roundtrip 
between Alice and Bob.
Because we assume coherent state light in our classical clock 
synchronization protocol, no loss
of generality ensues from this assumption.  In essence, any 
propagation loss in an actual
implementation can be accounted for by attenuating the input state 
used in the analysis below.

\begin{figure}[hbt]
\begin{center}\epsfxsize=.6
\hsize\leavevmode\epsffile{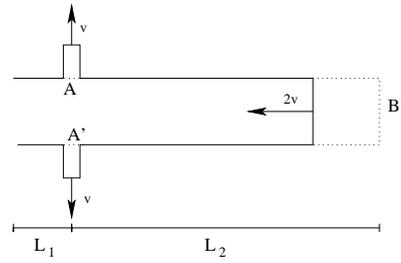}
\end{center}
\caption{Explanation of the delay $\tau_D$, from Eq.~(\ref{deftheta}),
that is due to the moving mirrors. Prior to the onset of mirror motion,
the total optical path length for the $\bquar$ polarization is
   $L=2L_1+2L_2$.  When the mirrors are moving, by the time the signal
   reaches point $A$, the first mirror has increased the path length by
$2v(L_1/c-t_0^a)$. This means that the $\bquar$-polarized signal will
incur a propagation delay $(L_1+L_2)/c+2v(L_1/c-t_0^a)/c$ en route to
point $B$. However, during this time interval, Bob's mirror has
reduced the path length for the $\bquar$ polarization by
$4v[(L_1+L_2)/c+2v(L_1/c-t_0^a)/c-t_0^b]$. Proceeding in a like manner
for the path length increase at $A'$, and
   summing up all the contributions, we can show that the overall delay
$\tau_D$ is given by Eq.~(\ref{deftheta}) to first order in $v/c$.}
\label{f:interf}\end{figure}

The $\tau_D$ expression in (\ref{deftheta}) is easily derived in the 
non-relativistic limit  $v\ll
c$ by observing that
${4v}(t_0^b-t_0^a)$ is the path length increase which the
interferometer introduces for the
$\bmquar$ polarization relative to the $\bquar$ polarization (see
Fig.~\ref{f:interf}).  In Eqs.~(\ref{delays1}) and (\ref{delays2}) the
terms
  \begin{eqnarray}
\kappa_\quar(\omega)&\equiv&\kappa_\quar^t(\omega)+
\kappa_\quar^f(\omega)\label{dispersive1}\\
\kappa_\mquar(\omega)&\equiv&\kappa_\mquar^t(\omega)+
\kappa_\mquar^f(\omega)
\;\label{dispersive}
\end{eqnarray}
represent the dispersive propagation medium encountered by
the $\bquar$ and $\bmquar$ polarizations;
$t$ refers to propagation {\it to} Bob, while $f$ refers to
propagation {\it from} him. We
neglected Doppler frequency shifts in deriving these dispersion terms;
see App.~\ref{s:relat} for a fully relativistic calculation.
Equations~(\ref{delays1}) and (\ref{delays2}) show that our
interferometer encodes the time-difference information
into both polarization components, whereas for synchronization
purposes it would be sufficient to encode such information on just
one.  [Thus, the scheme adopted here is an
instance of the differential conveyor belt protocol described in
App.~\ref{s:appa}]. However, as will be clarified later, the use of only
one polarization component does not provide dispersion immunity.

\begin{figure}[hbt]
\begin{center}\epsfxsize=.9
\hsize\leavevmode\epsffile{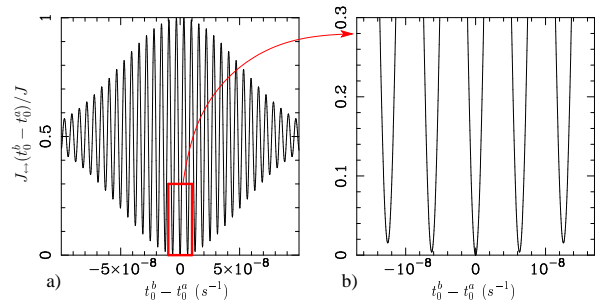}
\end{center}
\caption{a) Plot of $J_\leftrightarrow(t_0^b-t_0^a)$ versus $t_0^b-t_0^a$ from
   Eq. (\ref{totalint}) for Gaussian pulses. Here the velocity of the
   phase variation is $8v\omega_0/c=10^9$ s$^{-1}$ and the bandwidth is
   $\Delta\omega=10^{13}$ s$^{-1}$. b) Magnification of the box in the
   previous plot: $J_{\leftrightarrow}(t_0^b-t_0^a)$ has null at
   $t_0^a=t_0^b$.}
\label{f:dip}\end{figure}

The initial state of the system is a $\updownarrow$-polarized
coherent-state light pulse. It can be described in the frequency domain as
a tensor product of monochromatic coherent states of the form
\begin{eqnarray}
|{\Psi}\rangle&\equiv&\bigotimes_\omega|\alpha(\omega)\rangle_\updownarrow
|0\rangle_\leftrightarrow\nonumber\\&=&
\bigotimes_\omega|{\alpha(\omega)}/{\sqrt{2}}
\rangle_\quar|{\alpha(\omega)}/{\sqrt{2}}\rangle_\mquar\;
\label{statoin},
\end{eqnarray}
where the ket subscripts refer to polarizations and
$|\alpha(\omega)\rangle$ is a coherent state of frequency $\omega$
with amplitude function $\alpha(\omega)$ that has center frequency
$\omega_0$ and bandwidth $\Delta\omega$, e.g., a Gaussian.
Using Eqs.~(\ref{trasf1})--(\ref{delays2}) to express the
$\leftrightarrow$-polarized output field in terms of the
$\bquar$-polarized and $\bmquar$-polarized input fields and then
employing (\ref{statoin}) we obtain
  \begin{eqnarray}
I_\leftrightarrow(t)\!&=&\!\Bigg|\int {\rm d}\omega
\;\alpha(\omega)\sin\Big(\omega\tau_D-
\frac{\kappa_\quar(\omega)-\kappa_\mquar(\omega)}2\Big)
\nonumber\\&&
\times\;
e^{-i\omega(t-\tau) +i(\kappa_\quar(\omega)+\kappa_\mquar(\omega))/2}
\Bigg|^2
\;\label{intensit},
\end{eqnarray}
for the average photon flux at the $D_{\leftrightarrow}$ detector.
Dispersion in the propagation medium enters this expression through sum
and difference terms, i.e., $\kappa_\quar(\omega)+\kappa_\mquar(\omega)$
and
$\kappa_\quar(\omega)-\kappa_\mquar(\omega)$. The sum term does not
contribute to the output of an integrating detector,
\begin{equation}
J_{\leftrightarrow} \equiv \int {\rm d}t\, I_{\leftrightarrow}(t)\;.
\end{equation}
To suppress the difference term---and hence achieve dispersion
immunity---the two polarization components must undergo the same
dispersion in their roundtrip propagation between Alice and Bob, viz.,
\begin{eqnarray}
\kappa_\quar(\omega)=\kappa_\mquar(\omega)
\;\label{condiz}.
\end{eqnarray}
Under this constraint, the average photon number satisfies
\begin{eqnarray} J_{\leftrightarrow}(t_0^b-t_0^a) =2\pi\!\int\! {\rm d}\omega\,
|\alpha(\omega)|^2\sin^2\!\left(4v\omega(t_0^b-t_0^a)/c\right)\!
\label{totalint},
\end{eqnarray}
where our notation emphasizes the fact that the average photon number 
depends on the offset
between Alice's clock and Bob's. As shown in Fig.~{\ref{f:dip}}, the 
average photon number
consists of an envelope of duration
  $\sim$$v\Delta\omega/c$ that is modulated by fringes of frequency 
$8v\omega_0/c$,which result
from interference between the $\bquar$ and
$\bmquar$ return pulses at the polarizing beam splitter.  The mean 
value of this average photon
number fringe pattern is $J/2$, where
  \begin{eqnarray}
J  \equiv\int\!\! {\rm d}t\left|\int {\rm d}\omega\;\alpha(\omega)\;e^{-i\omega
     t}\right|^2
=2\pi\!\int\!\! {\rm d}\omega\;\left|\alpha(\omega)\right|^2
\label{intens}
\end{eqnarray}
is the average photon number of the input state (\ref{statoin}). 
(Remember that propagation loss
is ignored in our treatment.)  The extent of the fringe pattern is 
set by the clock offset
$|t_0^b-t_0^a|$ beyond which the $\bquar$ and $\bmquar$ return pulses 
do not overlap at the
polarizing beam splitter, so that no interference occurs.  When 
$t_0^b-t_0^a=0$, the average
photon number  $J_{\leftrightarrow}(t_0^b-t_0^a)$ vanishes, because 
the $\bquar$ and $\bmquar$
return pulses then arrive in synchrony and in phase, forming a 
$\updownarrow$-polarized field at
the polarizing beam splitter.   If we include propagation loss, then 
the occurrence of a perfect
$J_\leftrightarrow(0)$ null requires that the $\bquar$ and $\bmquar$ 
pulses encounter the same
loss in their roundtrip travel between Alice and Bob.  Such will be 
the case if: (a) we model loss
by assigning imaginary components to the dispersions $\kappa_\quar(\omega)$ and
$\kappa_\mquar(\omega)$; and (b) we require that (\ref{condiz}) be 
satisfied for the resulting
complex-valued dispersions.

Alice completes the conveyor-belt synchronization protocol by using a 
sequence of
pulses---shifted in time---to locate the null of the 
$J_\leftrightarrow$ interference pattern (see
Sec.~IIA for a more complete description).  The accuracy of such a 
measurement will be
$\sim$$c/v\omega_0\sqrt{\mbox{SNR}}$ , where $c/v\omega_0$ is the 
fringe width, and SNR is the
measurement signal-to-noise ratio that is achieved with this pulse 
sequence.  When $\mbox{SNR}\gg
1$, this accuracy can become comparable to the period $2\pi/\omega_0$ 
of the optical carrier
without violating our non-relativistic constraint, i.e., while 
maintaining $v\ll c$.  Note that
Alice can double the SNR of her synchronization by also observing the 
average photon number
$J_\updownarrow(t_0^b-t_0^a)$ from the $D_\updownarrow$ detector.  By 
energy conservation,
\begin{equation}
J_\updownarrow(t_0^b-t_0^a) + J_\leftrightarrow(t_0^b-t_0^a) = J,
\end{equation}
so that this additional measurement has a complementary fringe 
pattern, whose global maximum is
located at the offset between Alice's clock and Bob's.

In essence, our scheme embodies the precision of phase-locking schemes
such as {\cite{pharao}}, while maintaining the ability to directly
recover the time difference between Alice's clock and Bob's.  Interestingly,
because we measure the average photon number, i.e., the constant quantity
$J_{\leftrightarrow}(t_0^b-t_0^a)$, our protocol is immune to dispersion
provided that condition (\ref{condiz}) is satisfied. How can we enforce
such a condition in practice? Usually dispersion in an optical system 
is polarization
dependent,  so that Eq.~(\ref{condiz}) cannot be satisfied directly.
However, it is possible to transfer the polarization degree of freedom
to other degrees of freedom that undergo the same dispersion. For
example, if the medium is sufficiently homogeneous in space, then 
Alice may send her pulses as
co-polarized, spatially separated beams---which she recombines in an 
appropriate interferometer
after they return from Bob---to achieve the equivalent of 
(\ref{condiz}). Alternatively, if the
medium is sufficiently stable in time, then Alice may send two 
co-polarized, temporally separated
pulses that she recombines in a manner akin to the 
polarization-restoration scheme describe in
{\cite{jeff}} to achieve the equivalent of (\ref{condiz}).

\subsection{Multi-pulse  protocol}\label{s:step}
Alice needs to identify the global minimum---the null---of the 
$J_\leftrightarrow$ fringe pattern
in order to complete the conveyor-belt clock synchronization 
protocol.   In order to do so  she
will send a sequence of pulses, and employ the resulting 
$\leftrightarrow$ photon number
measurements from the $D_\leftrightarrow$ detector.  For each pulse, 
she will vary slightly the
delays that she imposes at points $A$ and $A'$, adding a distinct 
constant ${\cal T}_k$ to her
starting time $t_0^a$ for the $k$th pulse, viz., she will treat  the 
first pulse as if her clock's
initial time were  $t_0^a+{\cal T}_1$, she will treat the second 
pulse as if her clock's initial
time were $t_0^a+{\cal T}_2$, etc., something she can accomplish 
without knowing $t_0^a$.  Bob,
however, will continue to base his delays on the time shown on his 
clock.  From $\leftrightarrow$
photon number measurements made on this pulse sequence, Alice can 
estimate the fringe pattern
$J_\leftrightarrow(t_0^b-t_0^a)$, and hence pinpoint the location of the null.

\section{Quantum dispersion cancellation}\label{s:quantum}
The use of quantum resources can improve the performance of
traditional clock synchronization and positioning protocols
{\cite{paper}}.  The same is true for conveyor belt synchronization.
In particular, the use of frequency-entangled pulses offers greater 
immunity to dispersion than
is obtainable from the classical version of the protocol, as we now 
will show.  Suppose that the
input state to the Fig. {\ref{f:schema}} interferometer is  a stream 
of time-resolved,
frequency-entangled ($\omega_1 +
\omega_2 = 2\omega_0$) bi-photons from a type-II phase matched parametric
downconverter.   Instead of measuring the photon number at the output of
the
$D_\leftrightarrow$ detector, we now detect photon coincidences, i.e.,
near-simultaneous arrivals of photons at the $D_\leftrightarrow$ and
$D_\updownarrow$ detectors.   It can then be shown that  condition
(\ref{condiz}) for dispersion-immune classical operation is replaced by
the following less stringent condition under which the quantum system is
not degraded by dispersion:
\begin{eqnarray}
&&\kappa_\quar(\omega_0+\omega)
+\kappa_\mquar(\omega_0-\omega)=\nonumber\\&&
\kappa_\quar(\omega_0-\omega)
+\kappa_\mquar(\omega_0+\omega)
\;\label{miracolo}.
\end{eqnarray}
Interestingly, Eq.~(\ref{miracolo}) does not require the two
polarization components to undergo the same dispersion: this effect
results from the quantum frequency-correlations of the two
photons {\cite{kwiat,sasha,sasha1}}. As shown in {\cite{kwiat}},
Eq.~(\ref{miracolo}) will be satisfied when the odd-order terms
in the Taylor-series expansions of
$\kappa_\quar(\omega)$ and $\kappa_\mquar(\omega)$ about $\omega_0$ are equal.
We now present the essentials of the quantum dispersion cancellation 
derivation.

The clock synchronization signature that we are seeking is embedded 
in the probability that  the
$D_\leftrightarrow$ and $D_\updownarrow$ detectors both register 
photons within a coincidence
interval whose duration $T_c$ greatly exceeds $1/\Delta\omega$, the 
reciprocal of the
downconverter's fluorescence bandwidth, while still being short 
enough that the probability of two
bi-photons being present in this time interval is negligible.  This 
probability can be calculated
by considering a bi-photon initial state of the following form:
\begin{eqnarray} |\Psi\rangle\equiv\int
{\rm d}\omega\;\phi(\omega)|\omega_0+\omega\rangle_\quar
|\omega_0-\omega\rangle_\mquar
\;\label{tbst},
\end{eqnarray}
where $\phi(\omega)$ is the state's spectral function versus detuning 
$\omega=0$
from frequency degeneracy, i.e., when both component photons are at 
the center frequency,
$\omega_0$. The coincidence probability is then given by
\begin{eqnarray}
\Pr(t_0^b-t_0^a)
= \int\!{\rm d}t\int_{t-T_c/2}^{t+T_c/2}\!{\rm d}t'\,p(t,t'),
\end{eqnarray}
where
\begin{eqnarray}
p(t,t') \propto
\langle\Psi|E^{(-)}_\leftrightarrow(t)
E^{(-)}_\updownarrow(t') E^{(+)}_\leftrightarrow(t)
E^{(+)}_\updownarrow(t')|\Psi\rangle
\label{coincid},
\end{eqnarray}
is the joint probability density for detectors $D_\leftrightarrow$ 
and $D_\updownarrow$ to
register photons at times $t$ and $t'$, respectively. Unlike the 
classical case considered
earlier, in which the clock synchronization signature appeared in a 
fringe pattern, the
coincidence probability $\Pr(t_0^b-t_0^a)$ exhibits a ``Mandel dip'' 
(quantum interference)
{\cite{manou}} of width $\Delta\omega^{-1}$ whose null location is 
specified by the offset between
Alice's clock and Bob's:
\begin{eqnarray}
\lefteqn{P(t_0^b-t_0^a)\propto}\nonumber
\\
&&\hspace*{-.1in}\int\! {\rm d}\omega\,
|\phi(\omega-\omega_0)|^2\sin^2
\left(4v(\omega-\omega_0)(t_0^b-t_0^a)/c\right).
\end{eqnarray}
Thus, Alice can perform quantum dispersion-cancelling clock 
synchronization by a time-shifting
procedure similar to what we outlined in Sec.~\ref{s:step} for the 
classical case, obtaining an
accuracy $\sim$$1/\Delta\omega\sqrt{\mbox{SNR}}$.  An analogous 
quantum dispersion-cancelling
synchronization result was reported in {\cite{prl}}, using a 
different interferometer.

For the same SNR value, the classical synchronization system will 
outperform the quantum
synchronization system when $v/c > \Delta\omega/\omega_0$, a 
condition that is unlikely to be
satisfied for typical $\sim$THz downconverter bandwidths.  On the 
other hand, we may well inquire
whether a frequency-$\omega_0$ fringe pattern might be imposed onto 
the quantum system's Mandel
dip, dramatically enhancing its accuracy. From {\cite{sasha}} it 
appears that certain experimental
configurations allow fringes to be retained with use of the bi-photon 
state (\ref{tbst}). As the
authors of {\cite{sasha}} point out, however, there is no quantum 
dispersion cancellation in the
regime in which the fringes are present, {i.e.}, when the variable 
delay in their experiment is
placed {\it after} the beam splitter. In fact, it can  be shown that 
this regime does not exploit
the quantum correlation which is present in the state (\ref{tbst}): 
the signal from one of the
two detectors is used only to `filter out' a single
$\updownarrow$ polarized photon from the state (\ref{tbst}) which is
then sent into the interferometer. This means that the 
fringes-present regime in \cite{sasha}
is equivalent to a single-photon interferometer. So, had the authors 
of \cite{sasha}
measured the average photon flux resulting from a coherent-state 
input---instead of the
coincidences resulting from a bi-photon input---they would have 
obtained the same fringes.

\section{Conclusions}\label{s:concl}
We have presented an optical implementation of the ``conveyor belt'' clock
synchronization protocol that uses classical sources and, under rather
general conditions, is not disturbed by the presence of a dispersive
medium. The advantages of using quantum sources have been discussed
and compared with previous results on the same topic {\cite{prl}}.

\acknowledgments

We acknowledge A. V. Sergienko for interesting
comments and suggestions on quantum dispersion cancellation.  This
work has been supported by the ARDA, NRO, NSF, and by ARO under a MURI
program.

\appendix\section{}\label{s:appa}
In this appendix we discuss ways to
relax some of the requirements, described in Sec.~\ref{s:schema}, for 
the conveyor belt
synchronization protocol .

\begin{figure}[hbt]
\begin{center}\epsfxsize=.6
\hsize\leavevmode\epsffile{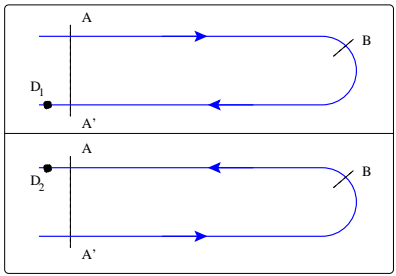}
\end{center}
\caption{Differential conveyor belt scheme: Bob is not required to be
   at the midpoint of the transmission line.}
\label{f:diff}\end{figure}

\paragraph*{Differential conveyor belt.---}

So far, we have assumed that the Alice-to-Bob and Bob-to-Alice 
propagation times are identical,
viz., $T_{ab} = T_{ba} = T$.  This amounts to having Bob located at 
the midpoint of the conveyor
belt in Fig.~\ref{f:nastro}.  We can eliminate this constraint by 
means of a differential
version of our protocol. Differential schemes---such as the two-way 
method for Einstein clock
synchronization---are conventionally employed to get rid of 
asymmetries.  The strategy we choose is
to introduce a second conveyor belt that proceeds in the opposite 
direction with respect to the
first one (i.e., it runs from $A'$ to $A$), as shown in 
Fig.~\ref{f:diff}. The protocol is carried
out as before: Alice and Bob respectively add and remove sand at 
points $A$, $A'$ and $B$, but
now they do this on both conveyor belts.  After the initial transient 
is over, the amount of
sand that Alice measures at the output of the first conveyor belt 
(point $D_1$ in
Fig.~\ref{f:diff}) is given by
\begin{eqnarray}
Q_{D_1}&=&\frac {s}{2}(t-T-T'-t_0^a)-s(t-T'-t_0^b)\nonumber \\ &+&
\frac{s}{2}(t-t_0^a) = s(t_0^b-t_0^a+T'-T)\;\label{qd1},
\end{eqnarray}
where $T$ is the transit time from $A$ to $B$ and $T'$ is the transit time
from $B$ to  $A'$.  Likewise, the amount of sand that Alice measures, after the
initial  transient, at point $D_2$ at the output of the second 
conveyor belt satisfies
\begin{eqnarray}
Q_{D_2}&=&\frac {s}{2}(t-T'-T-t_0^a)-s(t-T-t_0^b) \nonumber \\ &+&
\frac{s}{2}(t-t_0^a) = s(t_0^b-t_0^a+T-T')
\;\label{qd2}.
\end{eqnarray}
Clearly,
\begin{equation}
Q_{D_1}+Q_{D_2} = 2s(t_0^b-t_0^a),
\end{equation}
provides the desired synchronization information without requiring $T= T'$.

Note that the differential scheme requires that the forward
transmission times from $A$ to $B$ and from $B$ to $A'$ equal the
backward transmission times from $B$ to $A$ and $A'$ to $B$, 
respectively.  These equalities
can be achieved in optical implementations in which the forward 
(backward) transmitter and
backward (forward) receiver at $A$ ($A'$) are co-located.

\paragraph*{Imperfect clocks.---} The requirement that Alice and Bob
possess perfect clocks---i.e, that their clocks run at the same rate 
and do not drift appreciably
during a signal roundtrip time---may also be softened.  To do so, 
Alice must monitor the amount
of sand on the conveyor belt as a function of time, since it will not be a
constant, even after the initial transient has passed.  For example, 
suppose Alice and Bob have
drift-free clocks that run at different rates.  Insofar as the 
conveyor belt protocol is
concerned, this is equivalent to saying that Alice and Bob have 
clock's running at the same rate,
but that Bob uses  proportionality constant  $s'$, instead of $s$, 
when he removes sand from
point $B$.    Equation~(\ref{sabbia}) then becomes
\begin{eqnarray}
\!\!\!Q_D\!\!&=&\!\!\frac {s}{2}(t-2T-t_0^a)-s'(t-T-t_0^b)+
\frac{s}{2}(t-t_0^a) \nonumber\\&=&\!\!
(s-s')(t-T)+s't_0^b-s\;t_0^a
\;\label{sabbia4}.
\end{eqnarray}
Alice can now use a feedback loop to null out the $t$-dependent part 
of (\ref{sabbia4}) and thus
make her proportionality constant, hence her clock rate, the same as 
Bob's.   A similar procedure
will also work if Bob's clock drifts slowly---with respect to the 
signal roundtrip time---with
respect to Alice's.

\begin{figure}[t!]
\begin{center}\epsfxsize=.4
\hsize\leavevmode\epsffile{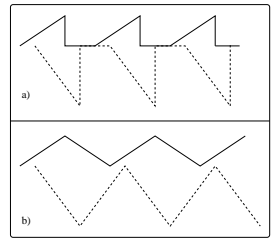}
\end{center}
\caption{Two examples of the periodic-ramps protocol. The lines
  plot the amounts of sand that Alice (solid) and Bob
   (dashed) must move to ($>0$) or from ($<0$) the conveyor belt 
versus time:  a)~Alice and Bob
   periodically restart the protocol; b)~Alice and Bob periodically
   reverse their rates.}
\label{f:rate}\end{figure}

\paragraph*{Periodic ramps.---} The conveyor belt protocol
requires Alice to deposit sand at rate $st^a/2$ and Bob to remove 
sand  at rate $st^b$.
With the passage of time, these requirements will soon get out of 
hand.  The essential behavior of
the conveyor belt protocol can be retained, however, by periodically 
restarting the protocol at
time intervals that are long compared to both the roundtrip 
propagation time and the offset
between Alice's clock and Bob's.  A more convenient alternative might 
be for Alice and Bob to
periodically reverse their rates, as shown in Fig.~\ref{f:rate}.  In fact, this 
periodic-ramp
approach is what is used in FMCW radar \cite{radar}.
\section{}\label{s:relat}
In this appendix we derive the relativistic corrections to
Eqs.~(\ref{deftheta})--(\ref{dispersive}). These corrections only 
matter if we violate $v/c\ll
1$.

We use Lorentz
transformations to go from the source outputs (in the laboratory 
reference frame), to the fields
at the moving mirrors (in the mirrors' reference frames), to the 
return pulses (back in the
laboratory reference frame), as described in {\cite{prl}}.  It is 
then possible to show that
(\ref{deftheta})-- (\ref{dispersive}) become
\begin{eqnarray}
\tau_D&\equiv&-\frac{4 v/c}{1-(v/c)^2}(t_0^b-t_0^a)
\;\label{deftheta1}\\
\kappa_\quar(\omega)&\equiv&\kappa_\quar^t(\omega/\chi)+
\kappa_\quar^f(\omega\chi)\\
\kappa_\mquar(\omega)&\equiv&\kappa_\mquar^t(\omega\chi)+
\kappa_\mquar^f(\omega/\chi),
\end{eqnarray}
where $\chi\equiv(1+v/c)/(1-v/c)$.
Moreover, a relativistic correction must also be applied to the
delay $\tau$ appearing in  Eqs.~(\ref{delays1}) and (\ref{delays2}):
\begin{eqnarray}
\tau\equiv \frac{2L}c\left(\frac{1+(v/c)^2}{1-(v/c)^2}\right)\;.
\end{eqnarray}

\end{document}